%% file: cosine-surfacealpha_v1.5.tex
\documentclass[preprint,5p,twocolumn]{elsarticle}

\def\rntt{$^{222}$Rn~}
\def\rasix{$^{226}$Ra~}
\def\pbten{$^{210}$Pb~}

\def\poten{$^{210}$Po~}

\usepackage{amssymb}

\usepackage[utf8]{inputenc}
\usepackage{stfloats}
\usepackage{lipsum}
\usepackage{amsmath}
\usepackage{xcolor}

\journal{Astroparticle Physics}

\begin{document}

\begin{frontmatter}



\title{Depth profile study of $^{210}$Pb in the surface of an NaI(Tl) crystal}

%
\author[a]{G.~H.~Yu}
\author[b]{C.~Ha}
\author[b]{E.~J.~Jeon\corref{cor1}}  
\ead{ejjeon@ibs.re.kr}
\author[b]{K.~W.~Kim\corref{cor1}}
\ead{kwkim@ibs.re.kr}
\author[b]{N.Y.~Kim}
\author[b,c,d]{Y.~D.~Kim}  
\author[b,d]{H.~S.~Lee} 
\author[e]{H.K.~Park}
\author[a]{C.~Rott} 
\cortext[cor1]{Corresponding authors}
\address[a]{Department of Physics, Sungkyunkwan University, Suwon 16419, South Korea}   
\address[b]{Center for Underground Physics, Institute for Basic Science~(IBS), Daejeon 34126, South Korea} 
\address[c]{Department of Physics, Sejong University, Seoul 05006, South Korea}
\address[d]{IBS School, University of Science and Technology (UST), Daejeon 34113, South Korea} 
\address[e]{Department of Accelerator Science, Korea University, Sejong 30019, South Korea}

\begin{abstract}
The surface $^{210}$Pb is one of the main background sources for dark-matter-search experiments using NaI(Tl) crystals,  
and its spectral features associated with the beta-decay events for energies less than 60~keV depends on the depth distribution of $^{210}$Pb in the surface of an NaI(Tl) crystal. Therefore, we must understand the profile of surface $^{210}$Pb to precisely model the background measurement in the low-energy region for the low-background experiment using NaI(Tl) crystals.  
We estimate the depth profile of the surface $^{210}$Pb contamination by modeling the measured spectrum of the alpha emission from the decay of $^{210}$Po at the decay sequence of the surface $^{210}$Pb contamination that is obtained using an $^{222}$Rn-contaminated crystal. 
In order to describe the energy spectra of the surface contamination we perform a log-likelihood fit of the measured data to a sum of Geant4 Mote Carlo simulations, weighted by an exponential curve as a function of the surface depth.
The low- and high-energy events from the beta decay of surface \pbten are also modeled to improve the depth profile for shallow depths. We simulate the energy spectra from beta decays of \pbten that are exponentially distributed in the surface by following two exponential functions 
where the mean-depth coefficients are free parameters in the data fitting; 
we observed that the energy spectra are in good agreement with the measured data.
\end{abstract}


\end{frontmatter}


\section{Introduction}
\label{intro}
The presence of dark-matter particles in the Universe has been evidenced via numerous astronomical observations~\cite{Komatsu:2010fb,Ade:2013zuv}. 
Weakly interacting massive particles (WIMPs) are one of the most attractive candidates for being dark-matter particles~\cite{lee77,jungman96}.
Many experiments have been conducted for directly searching WIMPs in our galaxy by looking for nuclear recoils that are produced by WIMP--nucleus scattering~\cite{gaitskell04,baudis2012}; however null results have been reported thus far  
with a notable exception, 
the DAMA/LIBRA experiment, which has consistently reported the observation of an annual event-rate modulation in an array of NaI(Tl) crystal detectors; this modulation, with 
a
statistical significance of more than 12.9~$\sigma$~\cite{Bernabei:2013xsa,Bernabei:2018yyw}, could be interpreted as a dark-matter signal~\cite{dama2008-2013}.
Furthermore, there are several NaI(Tl)-crystal-based experiments~\cite{Adhikari:2019,deSouza:2016fxg,amare:2019,sabre:2019,Fushimi:2015sew,Angloher:2016} to test DAMA/LIBRA's observation of an annual event-rate modulation.

Searching an annual modulation signal requires the complete understanding of background sources and complete simulation that accurately models the background-energy spectra measured using the detector; therefore, background models, which are based on Monte Carlo simulations, have been built using the Geant4 toolkit~\cite{geant4}~\cite{kimsnaibg:2017, cosinebg, cosmogenics:2020, anais2019, anais2016, anais2012}.
In these models, it has been reported that the low-energy contribution from the beta-decay of \pbten in the surface of NaI(Tl) crystals is one of the dominant background sources; furthermore, it has been suggested that sources are attributed to the \rntt contamination that occurred anytime during the powder- and/or crystal-processing stages~\cite{kimsnaibg:2017, cosinebg, anais2019, anais2016}.

\begin{figure}[t]
   \begin{center}
   \includegraphics[width=0.45\textwidth]{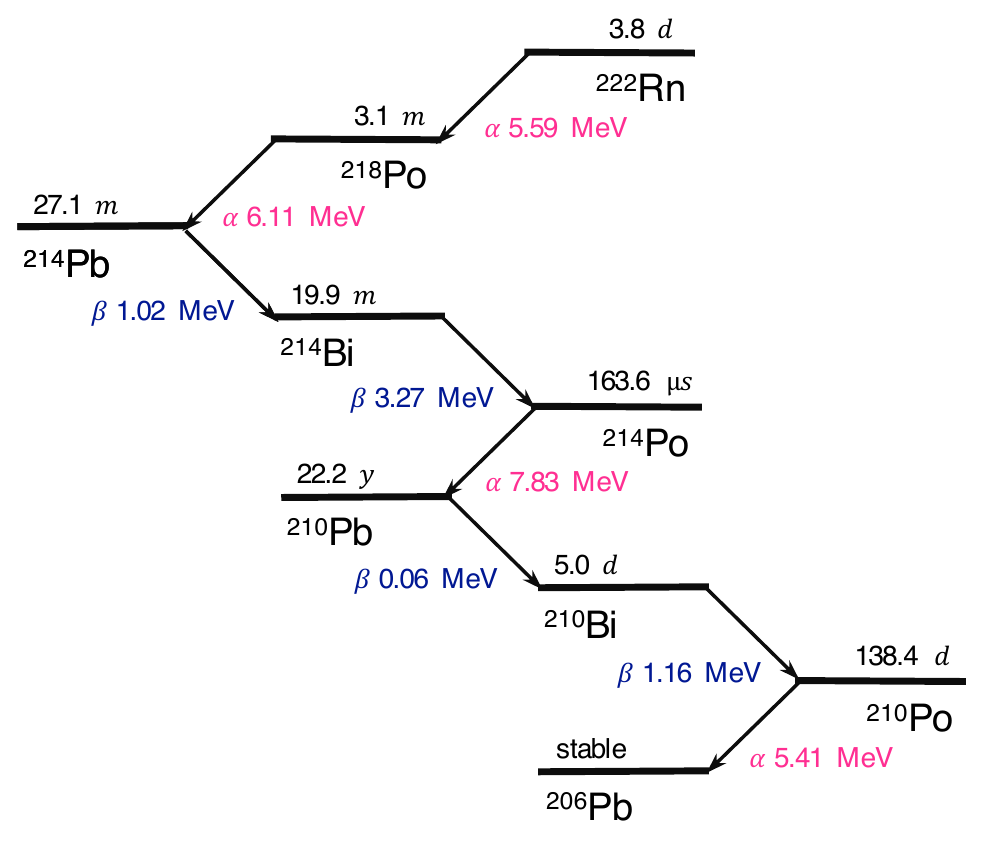}
   \caption{$^{222}$Rn decay chain. Alpha-decay Q values (down-left arrow), beta-decay Q values (down-right arrow), and  half-lives are indicated.
   }
   \label{decayscheme}
   \end{center}
\end{figure}

As depicted in Fig.~\ref{decayscheme}, the short-lived \rntt decays to $^{218}$Po, which can be deposited on the crystal surface as a reactive metal and recoils $^{214}$Pb into the crystal surface during its subsequent alpha decay~\cite{Smith:2000, Cooper:2000}. The $^{214}$Po alpha decay after the beta decays of $^{214}$Pb results in the implantation of \pbten deeper into the surface; consequently, \pbten has an implanted distribution in the surface and also a long half-life (t$_{1/2}$~=~22.3 years), thereby acting as a background source for the low-energy region in an NaI(Tl) crystal; in addition, it contributes to the low-energy spectra by the beta
decays to $^{210}$Bi, which decays to $^{210}$Po and subsequently,
decays via alpha emission to $^{206}$Pb. 

Because the beta decay to $^{210}$Bi results in low-energy events via the emissions of electrons and $\gamma$/X-ray, the spectral features of these events for energies less than 60~keV depend on the depth distribution of \pbten within the crystal surface. Therefore, we must completely understand the surface $^{210}$Pb profile to precisely model the background measurement in the low-energy region for 
low-background experiments
using NaI(Tl) crystals.
Additionally, there are other experiments not using NaI(Tl) crystals that improved the techniques to reject the surface background~\cite{superCDMS:2013, superCDMS:2017,CRESST:2015}.

To estimate the depth profile of the surface \pbten contamination, we model the measured spectrum due to the alpha emission from the decay of surface \poten (t$_{1/2}$~=~138 days) at the decay sequence of surface $^{210}$Pb (t$_{1/2}$~=~22.3 years), by performing Monte Carlo simulations using the Geant4 version 10.4.p02 (Sect.~\ref{sec:3}). In addition, the low- and high-energy events from the beta decay of surface \pbten are modeled to improve the depth profile for shallow depths (see Sect.~\ref{sec:4}).  
%

\input{analysis1_v1.5.tex}
\input{analysis2_v1.5.tex}

\section{Conclusion} 
\label{sec:conc}
We have studied the depth profile of the surface \pbten contamination of an NaI(Tl) crystal by modeling the measured energy spectra with the simulated spectra using the Geant4 toolkit. 
We first fitted the alpha-decay events with the simulated spectrum from the alpha decay of \poten that is exponentially distributed within the crystal surface 
by following an exponential function under the presence of an inactive region;
subsequently, we showed that the fitted result was in good agreement with the alpha data.  
In addition, to improve the depth profile for shallow depths below 1~$\mu$m, we studied the energy spectrum from the beta decay of \pbten whose spectral features were attributed to $\gamma$/X-rays and conversion electrons that were mainly distributed 
within shallow depths.
Subsequently, we showed that all the fitted results of both the clean and contaminated crystals were in good agreement with the data for both the low- and high-energy events. 

Using this study, we can provide the quantitative understanding of the depth profile of the surface \pbten contamination 
in an NaI(Tl) crystal exposed to $^{222}$Rn.
Furthermore, our analysis method and results can be applied to NaI(Tl) experiments, in general, to describe the backgrounds in case of \pbten surface contamination. 
Since the low-energy spectral feature due to the beta decays of the surface \pbten contamination is primarily attributed to depth profiles of \pbten distributed within a shallow surface with the mean depth of 0.107~$\mu$m as well as a deep surface with mean depth of 1.39~$\mu$m, it is expected to be similar for depth profiles of the surface \pbten in different crystals.
However, it could be affected by the $^{222}$Rn exposure.
There is also the possibility of bulk \pbten contamination, which is expected to affect the low-energy spectrum and, thus, the \pbten contribution is estimated by modeling the background from bulk \pbten and surface $^{210}$Pb.


\section*{Acknowledgments}
We thank the Korea Hydro and Nuclear Power Company for providing underground laboratory space at Yangyang. 
This work was supported by the Institute for Basic Science (South Korea) under project code IBS-R016-A1.



\bibliographystyle{elsarticle-num} 





\end{document}

%% file: analysis1_v1.5.tex
\section{Experimental setup}
\label{sec:2}
\begin{figure}[t]
   \begin{center}
   \includegraphics[width=0.4\textwidth]{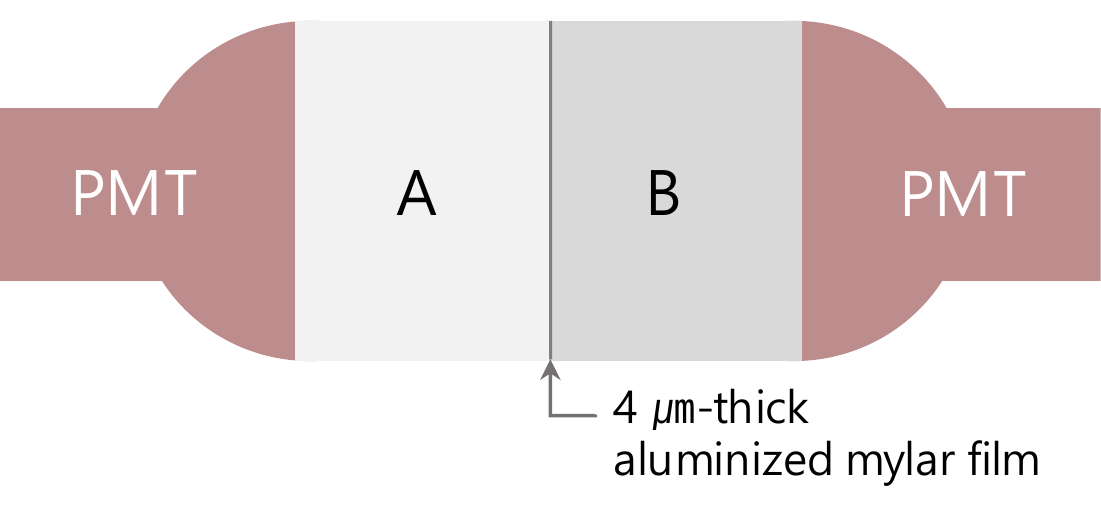}
   \caption{Schematic view of the detector. The crystal was cut into two pieces, and one of the pieces, i.e., Crystal B, was exposed to a radon source, while the other piece, i.e., Crystal A, was retained as clean. Subsequently, both the pieces were joined by inserting a 4-$\mu m$ thick mylar film, 
   which allowed alpha particles to pass through it but no optical photons.
   }
   \label{detector}
   \end{center}
\end{figure}

A copper encapsulated cylindrical NaI(Tl) crystal - with 8 cm diameter, 10 cm length, and 1.28 kg mass - was used in the experiment. 
It was made using the same ingot as the two crystals (C6 and C7) employed in the COSINE-100 experiment and 
encapsulated with the same copper that was used
for the COSINE-100 crystals~\cite{Adhikari:2017esn, Adhikari:2018ljm}.
The crystal was cut into two pieces, and the surface of one piece (Crystal B) was exposed to $^{222}$Rn gas from a $^{226}$Ra source for two weeks, while the other piece (Crystal A) was retained as clean.
Subsequently, both the pieces were attached facing each other  by inserting a 4-$\mu$m thick aluminized mylar film between them; the film not allowed scintillation photons of a few eV to pass through it. 
In addition, two photomultiplier tubes (PMTs), which detected the scintillation signals coming from the crystals, were attached at the ends of both the crystal.
The signal was saved only when the height of the pulse in both the PMTs was greater than the preset threshold. 
Figure~\ref{detector} depicts the crystal--PMT detector module, which was installed inside the CsI(Tl) crystal array setup  that had been previously deployed in the KIMS experiment in Yangyang underground laboratory~\cite{sckim12}.
Furthermore, Nitrogen gas was supplied into the detector setup to avoid radon contamination and maintain a stable humidity level;
the details of the experimental setup and data-recording conditions are described in Ref.~\cite{Kim:2018kbs}.

To minimize the contributions from the decay of the mother isotopes of \pbten and to ensure events primarily from the \pbten decay, only the data recorded after 70 days from the date of detector installation were used in this study; notably, the amount of data corresponds to 110 days.
The data were prepared using the following criteria.
An event that has hits only in the NaI(Tl) crystals~(Crystals A and B) is called a single-hit event, whereas the one that has accompanying hits in the surrounding CsI(Tl) crystals too is called a multiple-hit event.
For this study, we selected single-hit events since multiple-hit events are mostly due to external background sources.
In addition, a timing coincidence within 200 ns was required between Crystals A and B, called coincidence events, to consider the events that were mainly induced by the deposition on the crystal surface.
Furthermore, the energy scale below 100 keV was calibrated using a $\gamma$-ray source and the background spectra from several radioactivities of each crystal.
For Crystal A, we used the peaks at 12, 28, and 46.5~keV from the surface $^{210}$Pb, and at 59.54~keV from a $^{241}$Am $\gamma$-ray source. The peak at 76.63~keV contributed by X-rays from the decays of $^{214}$Pb and $^{214}$Bi was used for Crystal~B.
Above 100 keV, energy calibration was performed using peaks at 609 and 1120~keV from $^{214}$Bi  for both the NaI(Tl) crystals.

%% file: analysis2_v1.5.tex
\section{Modeling alpha spectra of $^{210}$Po decay in the crystal surface}
\label{sec:3}
 \subsection{Simulations of the surface $^{210}$Po contamination}
 \label{sec:3.1}
\begin{figure}[t]
   \centering
   \includegraphics[width=0.45\textwidth]{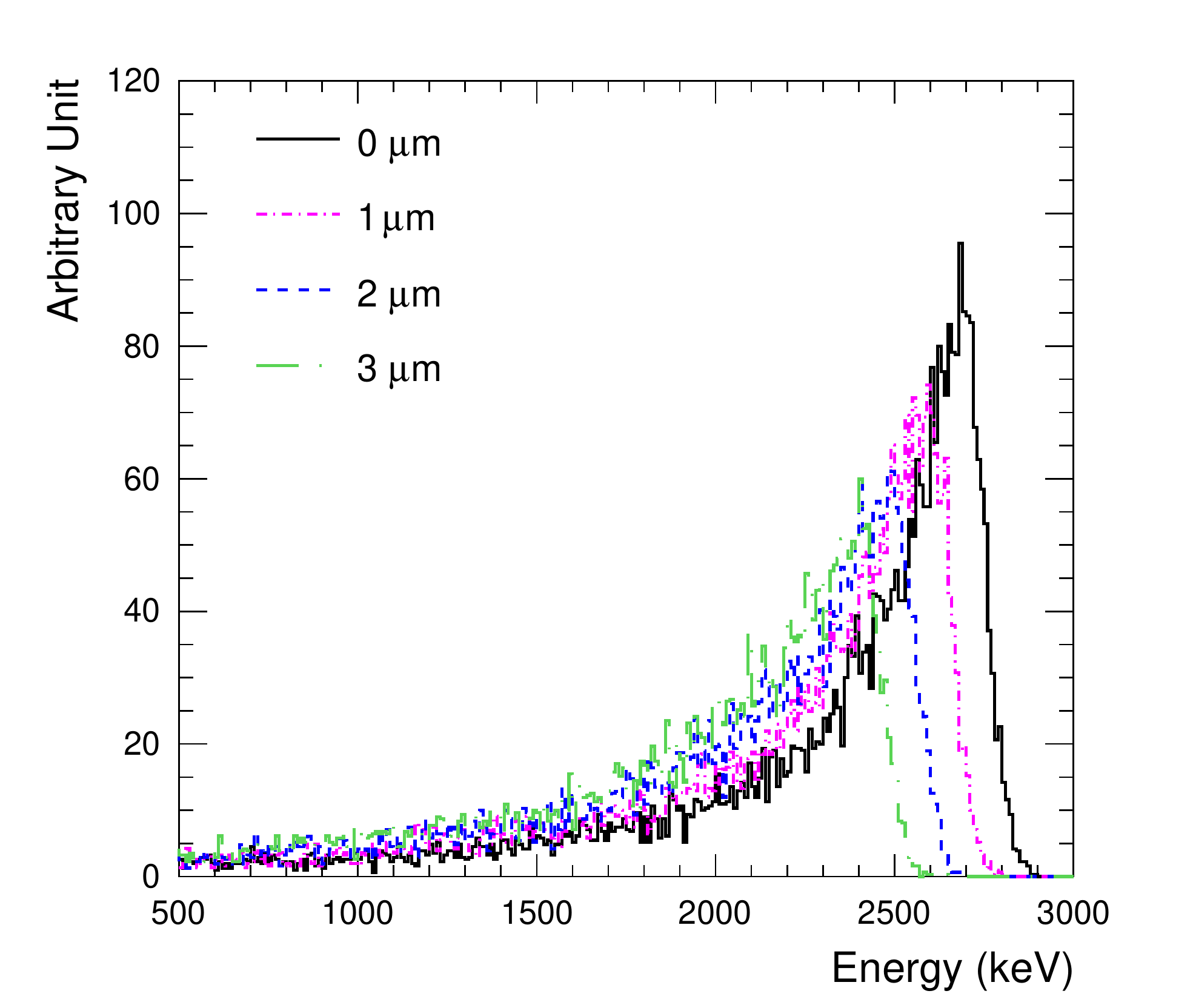} 
   \caption{
   Simulated alpha energy spectra deposited in Crystal A according to various depths of the surface of Crystal B.   
   We used the 5.3 MeV alpha energy from the decay of $^{210}$Po, considering 59\% of $\alpha$/$\beta$ light ratio at a given energy and assumed linearly quenched alphas, as determined by DAMA~\cite{bernabei2008}. 
   The property that the spectrum changes according to the depth position allows us to profile the depth-distribution with alpha spectrum.}
   \label{alphadepth}
   \end{figure}
    \begin{figure}[t]
    \centering
    \includegraphics[width=0.45\textwidth]{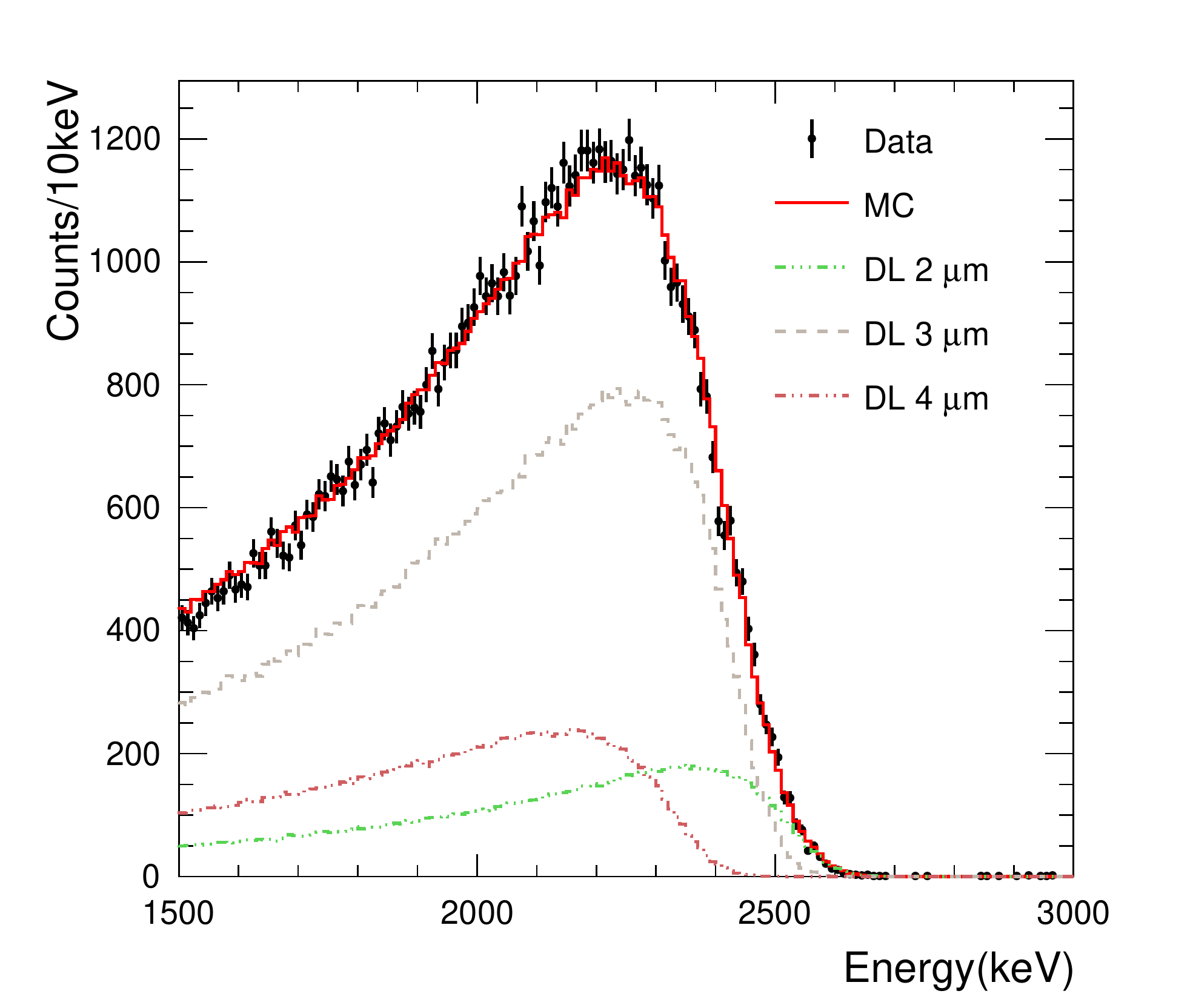} 
    \caption{
    Comparison of data and fitted result for the alpha spectrum on the clean Crystal A. 
    }
    \label{alpharesult}  
    \end{figure} 
As described in Sect.~\ref{sec:2}, the surface of Crystal B was exposed to \rntt gas from a \rasix source for two weeks; therefore, \pbten  at the decay sequence of \rntt exhibits an implanted distribution in the surface with a long half-life (t$_{1/2}$~=~22.3 years); it 
results in a low-energy spectra
because of the beta decays to $^{210}$Po, which subsequently decays via alpha emission to $^{206}$Pb. To estimate the depth profile of the surface \pbten  contamination, 
we modeled the spectrum measured by the clean crystal,
i.e., Crystal A; the spectrum is attributed to the alpha emission from the decay of surface \poten (t$_{1/2}$~=~138 days) of the contaminated Crystal B, by simulating all the decay chains of \pbten 
with origin in the surface of Crystal B.
We, thus, 
generated \pbten through different depths
of the surface of Crystal B, and investigated the correlation of alpha spectra measured using clean Crystal A in terms of the surface depth. 
Because of both the thin mylar layer and the small energy depositions on the contaminated crystal, i.e., Crystal B, the measured energy of the alpha particles was below the full energy deposition of $^{210}$Po, as reported in Ref.~\cite{Kim:2018kbs}; in addition,  their small energy depositions on Crystal B vary depending on the 
depth of the contamination,
thereby resulting in different alpha spectra on clean Crystal A, as depicted in Fig.~\ref{alphadepth}. 
These were used as inputs when fitting the measured data to the sum of these MC spectrums, weighted by an exponential curve as a function of the contamination depth;
the details are described in Sect.~\ref{sec:3.2}.

To compare the energy spectrum of the alpha particles from the simulations with the data, 
we used the 5.3 MeV alpha energy from the decay of \poten
that is measured to be 3.126$\pm$0.008~MeV electron equivalent energy in the clean Crystal A, 
corresponding to 59\% of light ratio at 5.3~MeV. This value is in good agreement with the quenching factor measured by DAMA at this energy~\cite{bernabei2008}, so in the following we assume the same energy dependence as in Ref.~\cite{bernabei2008}.
Upon contaminating a crystal by exposing it to \rntt gas, a foggy layer was formed on its surface; this layer can affect the scintillation performance by acting as an inactive region or a dead layer (DL)~\cite{yang2014}; therefore, the contaminated crystal, i.e., Crystal B, appeared obscured, while the clean Crystal A appeared transparent. We, thus, 
considered a DL of variable thickness
in the surface of contaminated Crystal B.

\subsection{Results and comparison with measured data} 
\label{sec:3.2}
To model the surface \pbten spectrum, we generated \pbten decays at random locations within the surface thickness of 10~$\mu$m in the contaminated Crystal B, which was divided into 100 bins, each with a thickness of 0.1~$\mu$m, to control the fraction of events on the basis of depth.  
We assume that they are exponentially distributed in the surface and the energy spectrum deposited in Crystal~A, $N(E)$, is as follows:
%
    \begin{equation}
	N(E) = p_{0}\sum_{DL=0}^{5}F_{DL}\sum_{d=0}^{10\mu m}N_{d}^{DL}(E)\exp\left(\frac{-d}{p_{1}}\right)
	\label{eq1}
    \end{equation}    
where $F_{DL}$ represents the fraction of the dead layer of thickness $DL$ $\mu$m, $d$ denotes the contamination depth in $\mu$m, $N_{d}^{DL}(E)$ the energy spectra deposited in Crystal A by the events generated at the surface depth $d$ for every $DL$, and $p_{0}$ and $p_{1}$ the amplitude and mean depth of the exponential distribution, respectively. 
%
Subsequently, we fitted the alpha spectrum measured using the clean Crystal A by employing the log likelihood method, which allows the simulated spectrum to follow Eq.~(\ref{eq1}), 
with free-floating parameters
$p_{0}$, $p_{1}$, and $F_{DL}$.
In the fit, we included five DLs (DL1, DL2, DL3, DL4, and DL5) of various thicknesses from 1 to 5~$\mu$m as follows: 0$\sim$1, 0$\sim$2, 0$\sim$3, 0$\sim$4, and 0$\sim$5~$\mu$m, and their fractions were treated as free-floating parameters. 
DL0 represents no dead layer.
Figure~\ref{alpharesult} depicts the fitted result (solid red line) for the alpha spectrum on the clean Crystal A; the result is in good agreement with the data (filled black circles). 
The result supports a \poten distribution following an exponential function with best fit parameters,
$p_{1}$~=~(1.39$\pm$0.02)~$\mu$m and the fractions of dead layers that are 0\%, 0\%, 5.5\%, 56.2\%, 38.2\%, and 0\% for DL0, DL1, DL2, DL3, DL4, and DL5, respectively. 
The area under the DL4 curve shown in Fig.~\ref{alpharesult} is slightly larger than that of DL2 while the fractions of DL2 and DL4 are resulted in 5.5\% and 38.2\% from the fit. It is because less events are tagged as coincidence events when they are simulated including a thick dead layer.
Accordingly, it shows that the alpha-decay events from \poten generated on the surface for the DLs with thicknesses of 3 and 4~$\mu$m are dominant contributors.

\section{Modeling the energy spectra of $^{210}$Pb beta decay within the crystal surface}
\label{sec:4}
	\begin{figure}[ht]
   	\begin{center}
   	\includegraphics[width=0.49\textwidth]{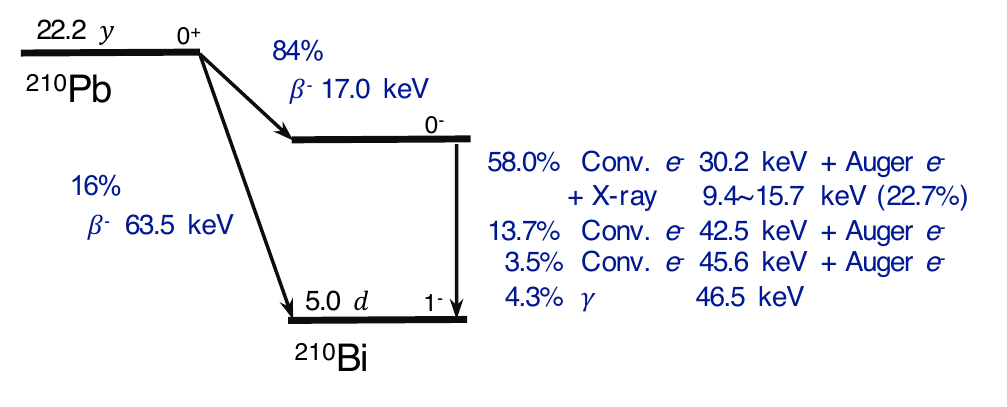}
   	\caption{Beta decays of $^{210}$Pb.}
   	\label{betadecayPb210}
   	\end{center}
	\end{figure}
        \begin{figure*}[ht]
        \centering
        \begin{tabular}{cc} 
        \includegraphics[width=0.45\textwidth]{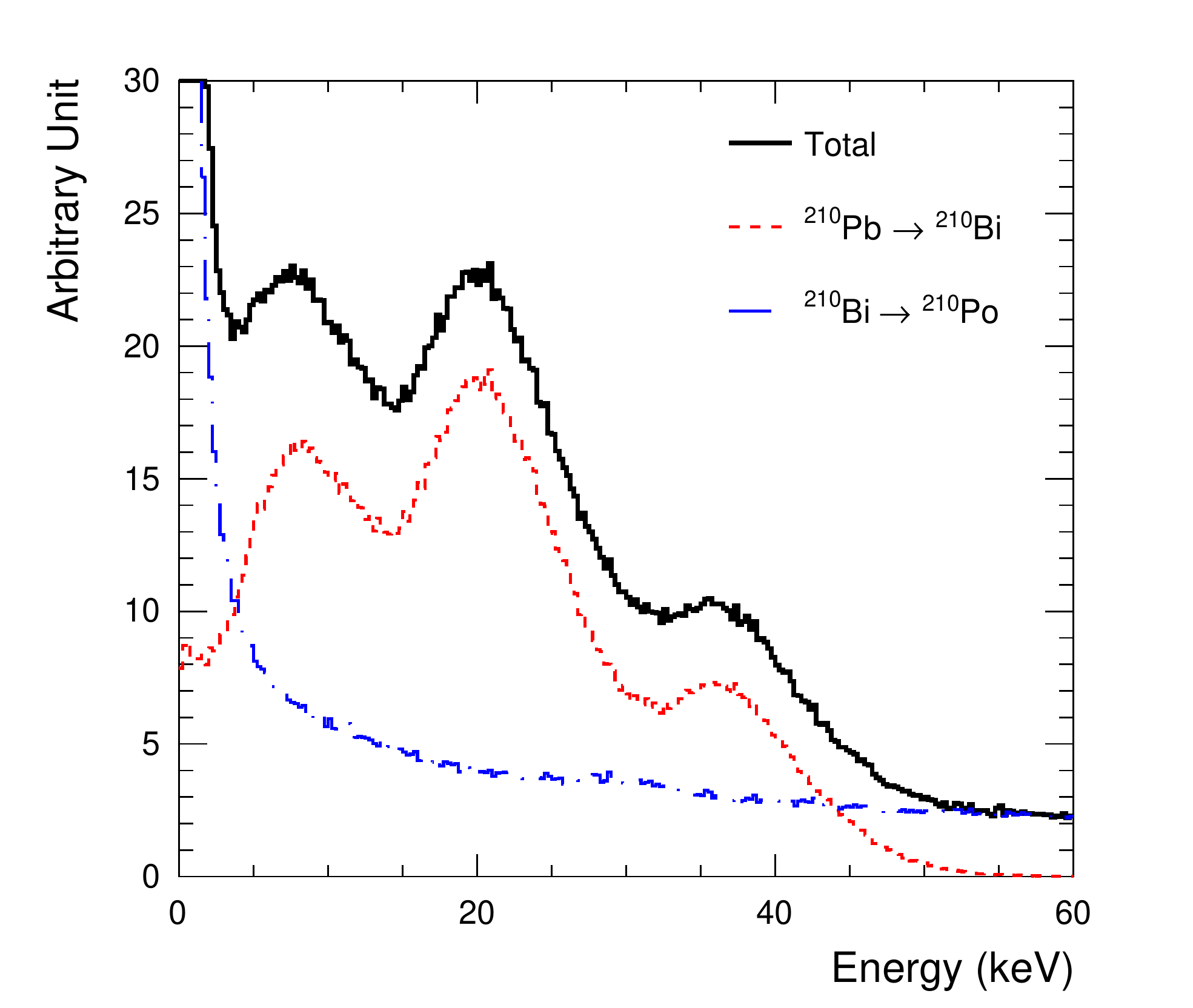} &
        \includegraphics[width=0.45\textwidth]{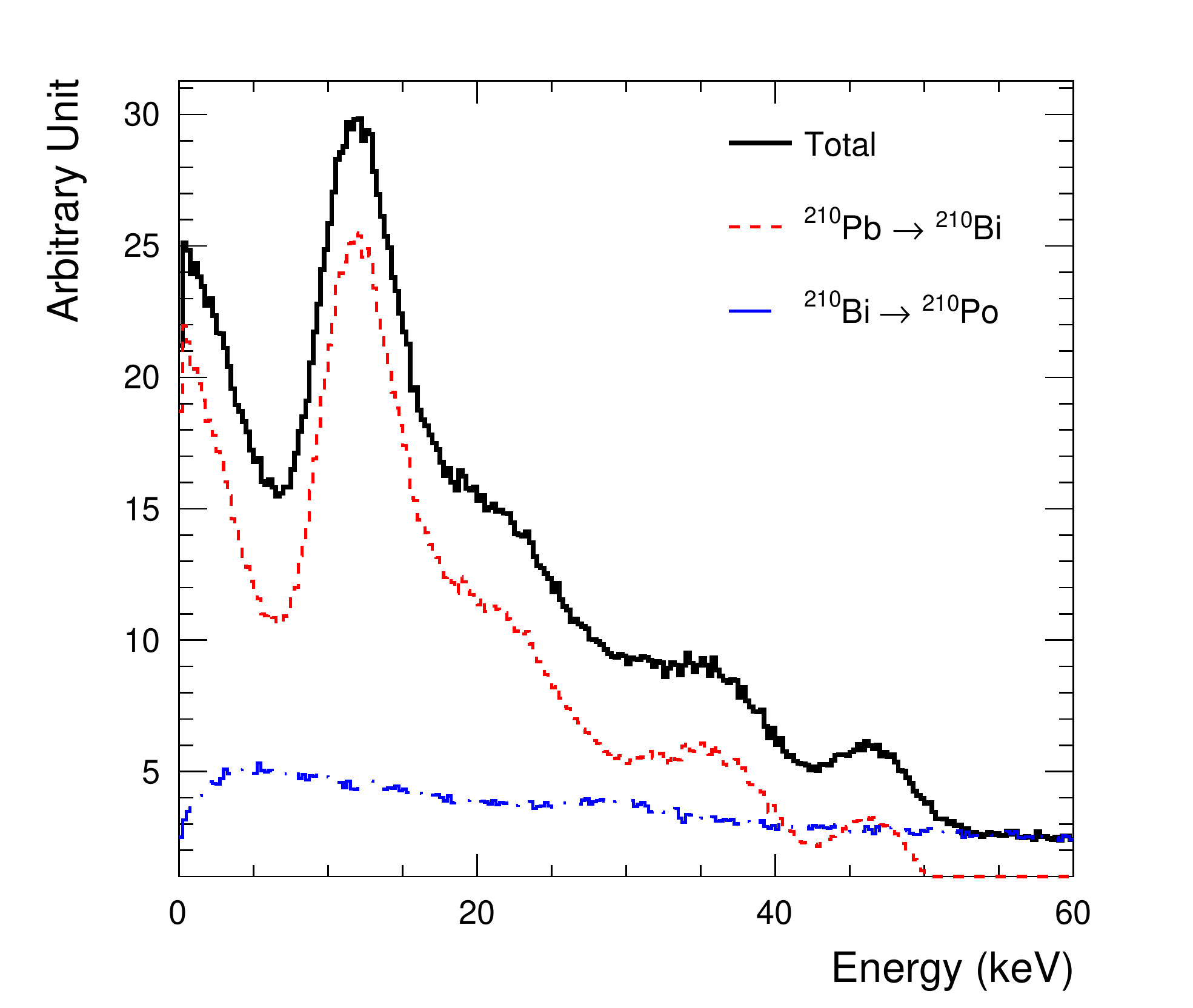} \\
        (a) & (b) \\
        \end{tabular}
        \caption{
        Low-energy spectra due to the beta decay of \pbten to $^{210}$Po: in the contaminated Crystal B (a), and in the clean Crystal A (b).}    
        \label{betadecay} 
        \end{figure*}
        \begin{figure}[!b]
        \centering
        \includegraphics[width=0.49\textwidth]{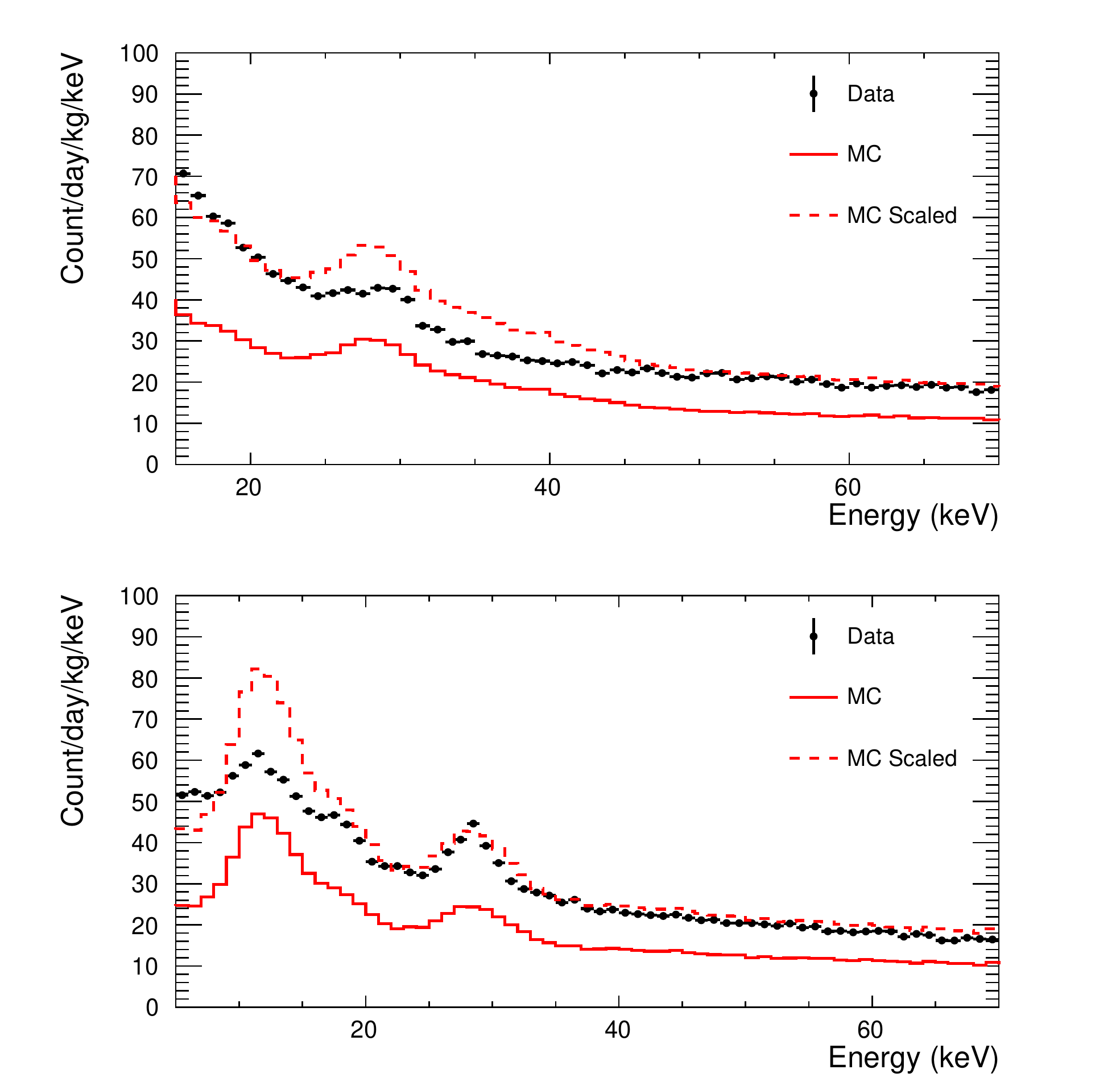}
        \caption{
        Low-energy spectra due to the beta decay of \pbten generated in the surface with the mean-depth parameter, $p_{1}$, 
including
        all the energy deposited in the clean Crystal A (bottom) and in the contaminated Crystal B (top). The energy spectra are compared with the measurement results (filled black circles).
Dashed line: scaled spectrum (see text).
	}
        \label{cleananode1}
        \end{figure}        
        \begin{figure*}[tbp]
        \centering
        \includegraphics[width=0.95\textwidth]{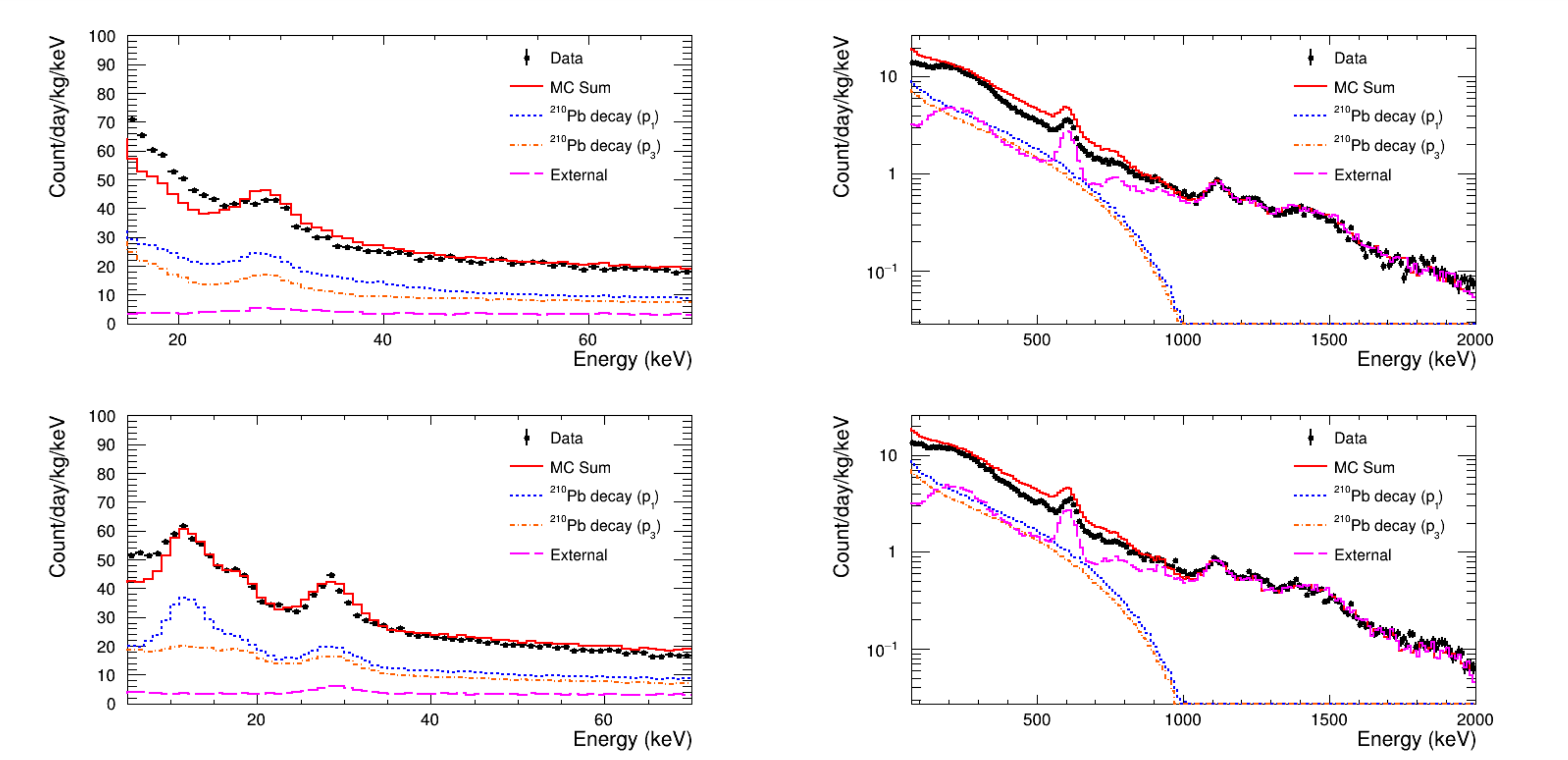}
        \caption{
	Fitted energy spectra for both Crystals A (bottom) and B (top) are compared with the measurement results (black circles). The left ones represent the low-energy spectra and the right ones the high-energy spectra of the beta-decay events.
	}
        \label{gammaresult}
        \end{figure*}   
        \begin{figure*}[tbp]
        \centering
        \begin{tabular}{cccc} 
        \includegraphics[width=0.22\textwidth]{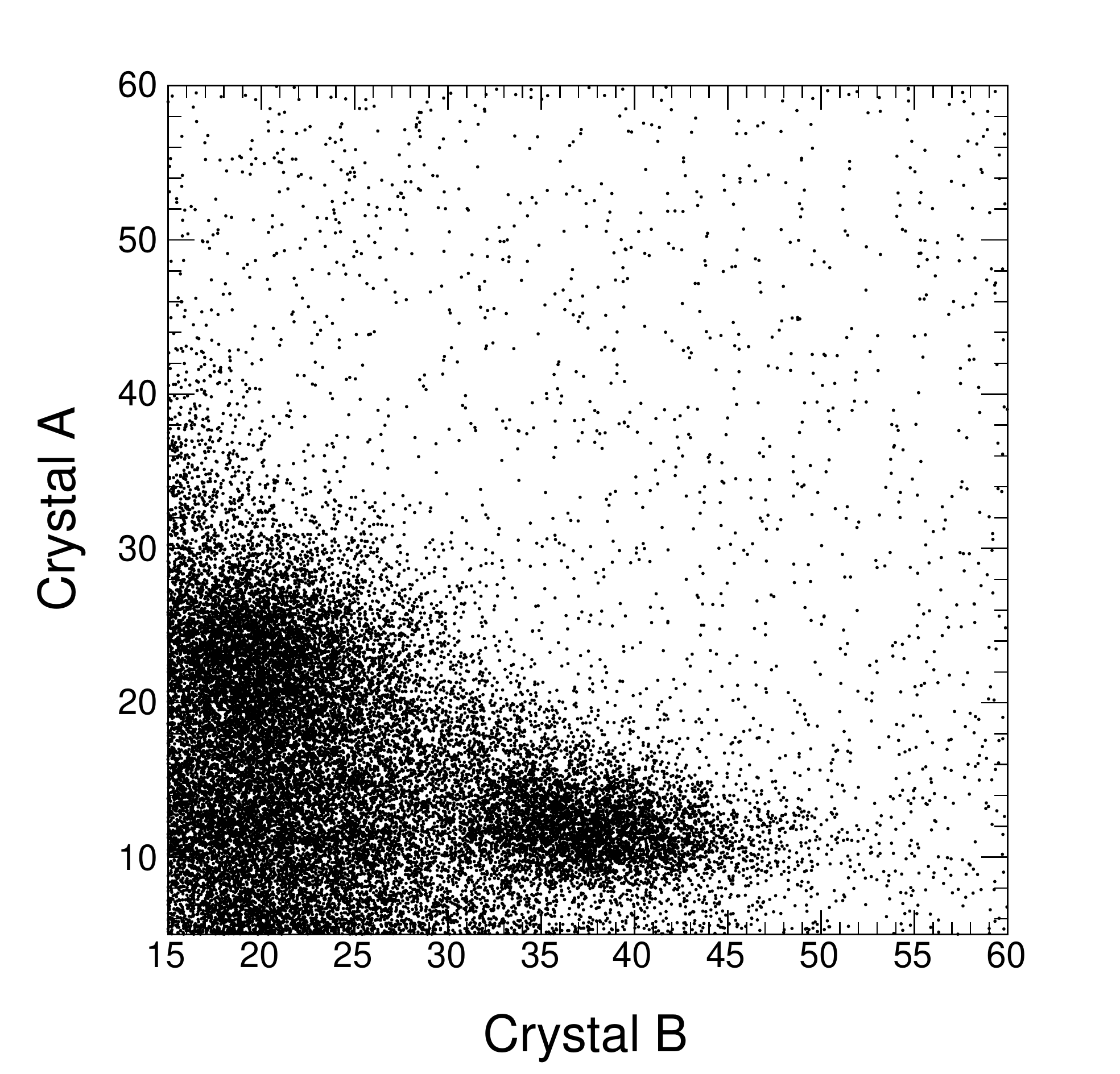} &
        \includegraphics[width=0.22\textwidth]{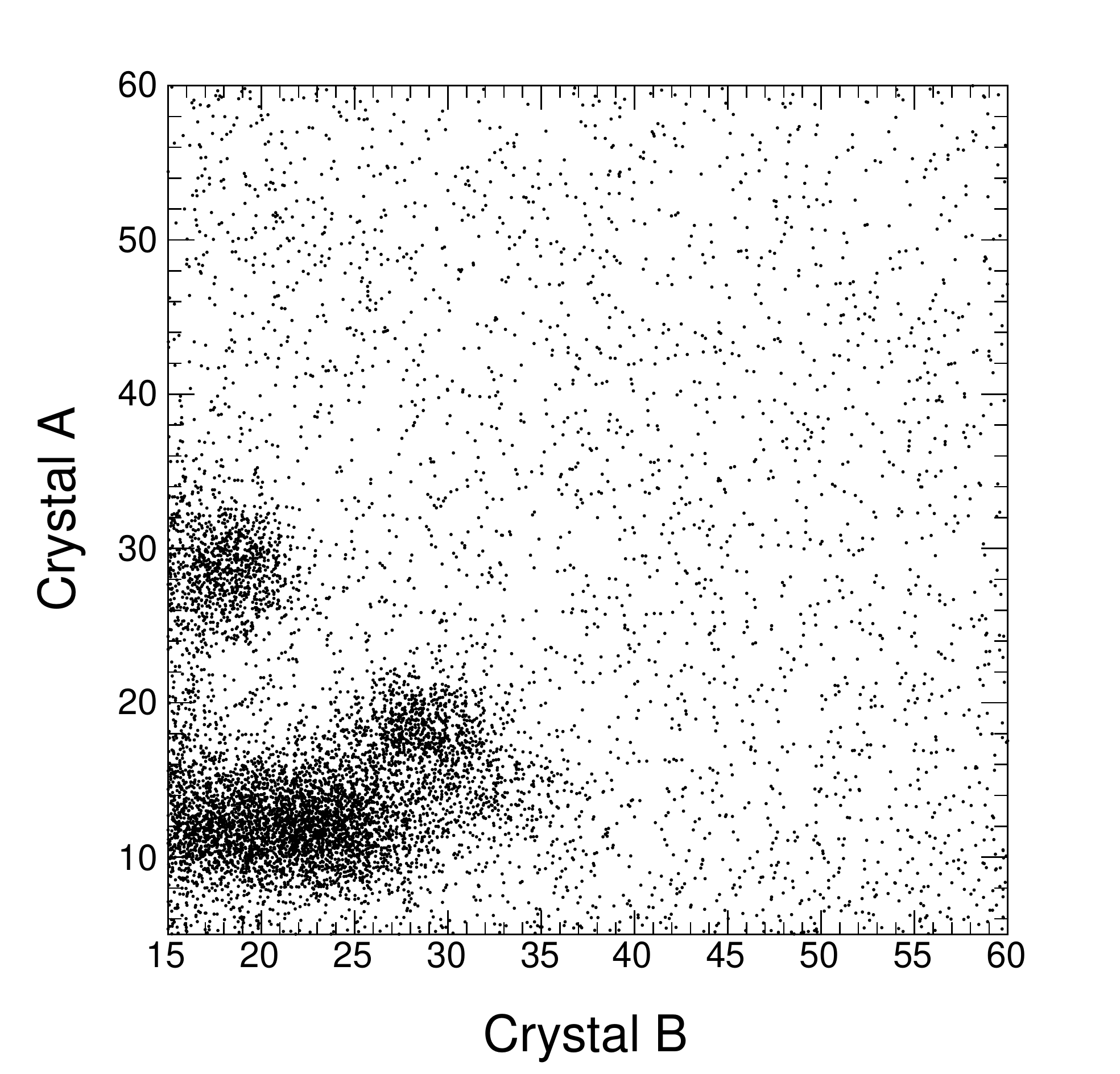} &
        \includegraphics[width=0.22\textwidth]{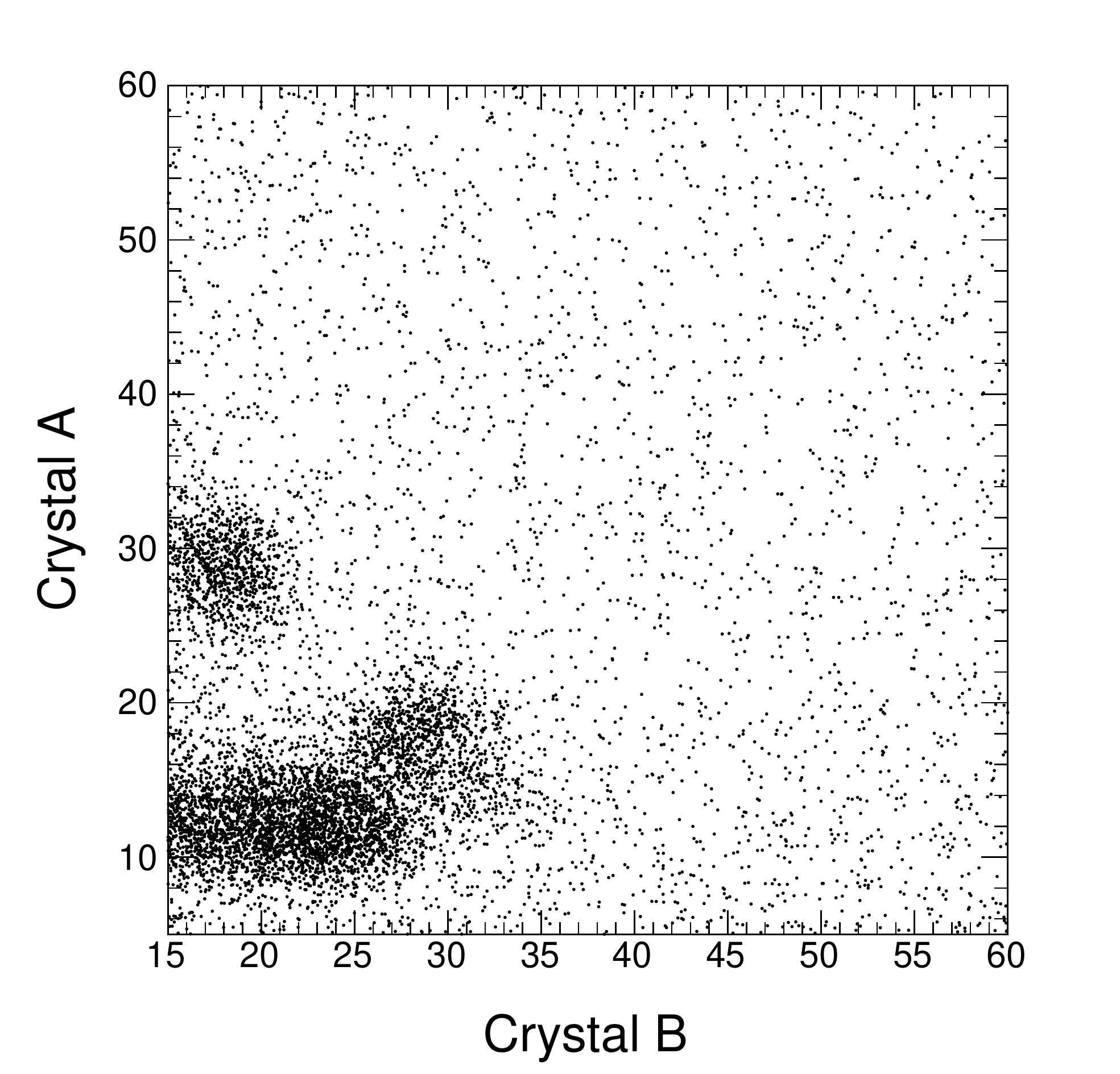} &
        \includegraphics[width=0.22\textwidth]{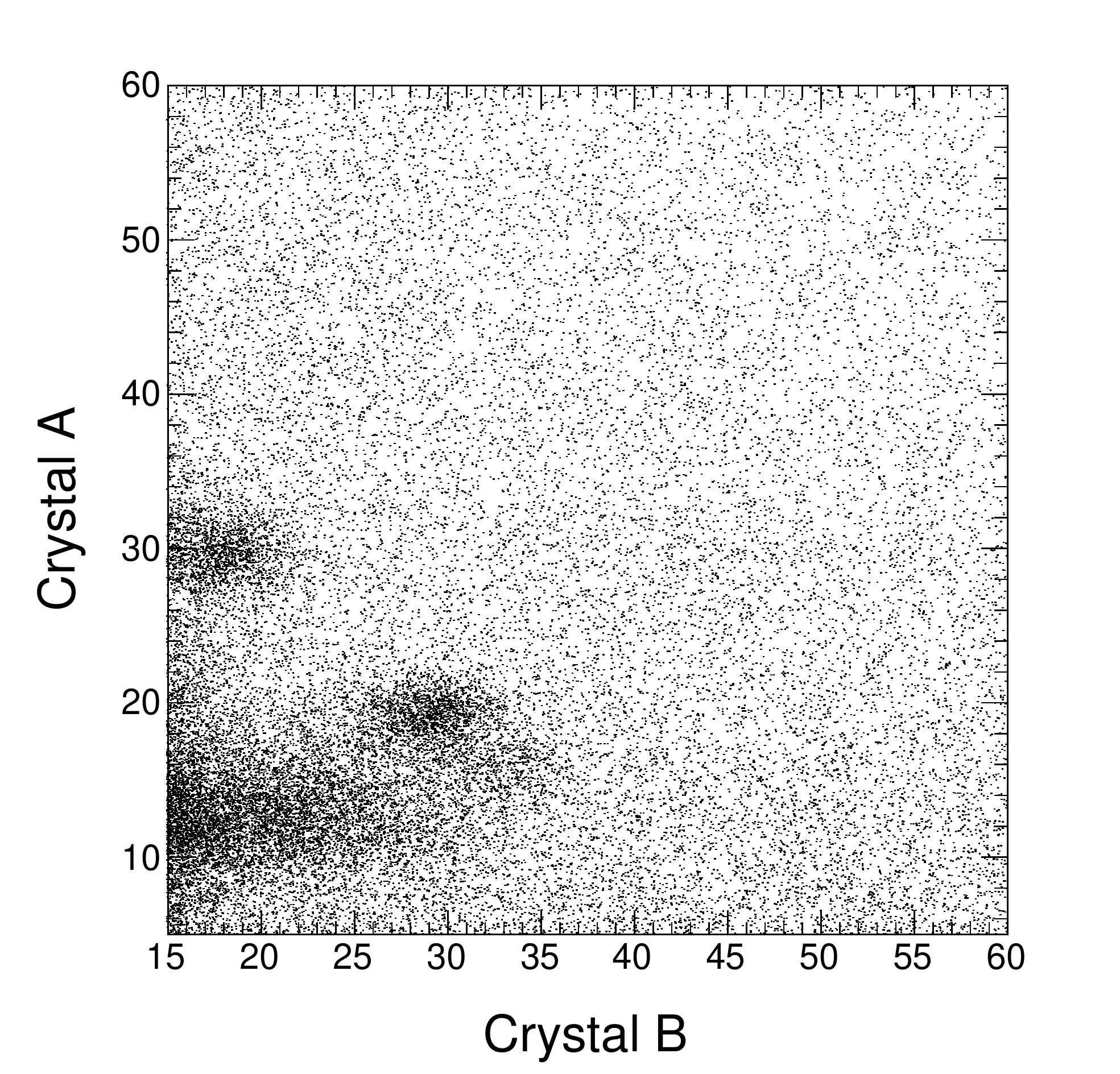} \\
        (a) & (b) & (c) & (d) \\
        \end{tabular}
        \caption{
        Energy deposited in Crystal A versus that deposited in Crystal B (in units of keV). (a), (b), and (c) represent the simulated results and (d) the measured data. 
        We generated the \pbten decays at random locations within the surface thickness of 0.5~$\mu$m in the contaminated Crystal B both without a DL (a), with a 3-$\mu$m thick DL (b), and with a 4-$\mu$m thick DL (c). 
       }    
        \label{scatterplot} 
        \end{figure*}

As shown in Fig.~\ref{betadecayPb210}, the beta decays to $^{210}$Bi from \pbten produced low-energy events via the emissions of electrons and $\gamma$/X-ray; therefore, their spectral features for energies less than 60~keV depend on the depth distribution of \pbten on the crystal surface.
To understand the energy spectra from the beta decays of $^{210}$Pb, we simulated them by generating \pbten at random locations within the surface thickness of 1~$\mu$m in Crystal B. The simulated spectra for both the contaminated Crystal B and clean Crystal A are depicted in Fig.~\ref{betadecay} (a) and (b), where each color represents the beta decays of \pbten (dotted red line) and $^{210}$Bi (dashed blue line), respectively. The peaks at approximately 10 and 46~keV are attributed to the X-rays and 46.5~keV $\gamma$-ray from the decays of $^{210}$Pb. In addition, the conversion electrons contribute to the peaks at approximately 20 and 35~keV. Therefore, these spectral features depend on the beta-decay events distributed within a shallow surface depth. 

To verify the depth profile derived using the alpha-spectrum model, which is described in Sect.~\ref{sec:3.2}, we compared the low-energy spectra measured using each of Crystals A and B with the simulated spectra of \pbten that is distributed in the surface of Crystal B, in accordance with the depth profile derived using the alpha spectrum; 
we observed that the low-energy spectra were not satisfactorily reproduced using the simulated spectra in the low-energy region, as depicted in Fig.~\ref{cleananode1} (solid red line). To avoid the randomly coincident events in the low energy region due to the backgrounds and noise in the $^{222}$Rn-contaminated Crystal B and Crystal A, we do not use those events below 5~keV in the clean Crystal A and 15~keV in the contaminated Crystal B, respectively.

If the \poten contamination is zero at the 
moment when the initial \pbten contamination occurred, it will grow with the \poten half-life, i.e., $\tau_{Po_{210}}$~=~138 days, until reaching an equilibrium. This change in the total alpha rate  with time can provide information regarding the total amount of the surface \pbten contamination. Because the 110~day data used in this study were recorded after 70~days from the time when the initial \pbten contamination occurred and when an equilibrium was not yet reached, the \poten contamination is lower than the surface \pbten contamination. However, the spectral feature of the simulation is not satisfactorily matched to the data, 
even when scaled,
as shown in the dotted red line of Fig.~\ref{cleananode1}.
The 
spectral feature
discrepancy between the data and simulation can be explained as follows: 
most alpha events occurring at the shallow depth of  less than 1~$\mu$m are not tagged as coincidence events when it is dominated by an inactive region and when the path length of the recoiled $^{206}$Pb with 106~keV kinetic energy is as small as $\sim$50~nm; therefore, it results in the lack of full understanding of the depth profile for shallow depths.

Therefore, to improve the model for better understanding the depth profile of \pbten for shallow depths, we assume that they  are exponentially distributed in the surface of Crystal B, via following two exponential functions:
%
    \begin{align}
	N(E) &= 
		\begin{aligned}[t]
		&\sum_{DL=0}^{5} F_{DL}\left[ p_{0}\sum_{d=0}^{10\mu m} N_{d}^{DL}(E) \exp\left(\frac{-d}{p_{1}}\right)\right. \\
		&\left.+~p_{2}\sum_{d=0}^{10\mu m} N_{d}^{DL}(E) \exp\left(\frac{-d}{p_{3}}\right)\right]
		\end{aligned}
    \end{align} 
where we constrain the parameters $p_{1}$ and $F_{DL}$  using (1.39$\pm$0.02)~$\mu$m and the fractions of dead layers obtained from the alpha spectrum modeling described in Sect.~\ref{sec:3.2} and treat $p_{0}$, $p_{2}$, and $p_{3}$ as free-floating parameters in the data fitting.
In the fit, the low- and high-energy spectra of the beta-decay events, as well as the energy deposited in each of Crystals A and B, are simultaneously fitted using simulations.
The fitted energy spectra for both Crystals A and B~(solid red line) are compared with the measurement results~(filled black circles), as depicted in Fig.~\ref{gammaresult}. The overall energy spectra (solid red lines) for both the clean Crystal A and contaminated Crystal B are in a good agreement with  the data, not only for low-energy events but also for high-energy events.
The high-energy events are dominantly the result of the external $^{226}$Ra from the copper cylinder that encapsulates the NaI(Tl) crystal 
that also slightly contributes to the low-energy region and beta decay from $^{210}$Bi to \poten in the crystal surface that ends with 1.2~MeV.   
Since we use the coincidence events there is no significant chance to the coincide in both crystals by the backgrounds from the PMTs that is the main contributor to the external background.
As a result of the fit, the mean of the exponential depth distribution with $p_{3}$ is as shallow as (0.107$\pm$0.003)~$\mu$m. 
The ratio of the amplitudes of two exponential distributions, $p_{0}$ to $p_{2}$, is approximately 0.76; therefore, the depth distribution of \pbten in the surface of an NaI(Tl) crystal consists of both shallow and deep depth profiles.

According to the decay chain for $^{210}$Pb, most events with energy deposition in the crystals are attributed to the conversion electrons, Auger electrons, and $\gamma$/X-rays, followed by the beta electrons from the decay to $^{210}$Bi of \pbten  in the surface of Crystal B. 
Therefore, the coincidence events might possess an energy correlation between Crystals A and B. 
In Fig.~\ref{scatterplot}, we depict a scatter plot of the energy deposition in Crystal A versus that in Crystal B. Figure~\ref{scatterplot}(a), (b),  and (c) depict the simulated results both without including a DL and upon including a DL, respectively; Fig.~\ref{scatterplot}(d) depicts the measured data. 
In the simulation, because it is found to be in good agreement with the data when it is dominated by 3 and 4-$\mu$m thick DLs in the fit, 
we included 3 and 4-$\mu$m thick DLs while generating \pbten decays at random locations within the surface thickness of 0.5~$\mu$m in the contaminated Crystal B; subsequently, we reproduced the distinct islands at approximately 20 and 30~keV in one crystal and at 30 and 20~keV in the other one, as evident from the measured data, while the islands are not depicted in Fig.~\ref{scatterplot}(a), which does not consider a DL.  